# Stacking-order effect on spin-orbit torque, spin-Hall magnetoresistance, and magnetic anisotropy in $Ni_{81}Fe_{19}$-$IrO_2$ bilayers

Kohei Ueda[1,2*], Naoki Moriuchi[1], Kenta Fukushima[1], Takanori Kida[3], Masayuki Hagiwara[3], and Jobu Matsuno[1,2]

[1]*Department of Physics, Graduate School of Science, Osaka University, Osaka 560-0043, Japan*
[2]*Center for Spintronics Research Network, Graduate School of Engineering Science, Osaka University, Osaka 560-8531, Japan*
[3]*Center for Advanced High Magnetic Field Science, Graduate School of Science, Osaka University, Osaka 560-0043, Japan*

The 5$d$ transition-metal oxides have been an intriguing platform to demonstrate efficient charge to spin current conversion due to a unique electronic structure dominated by strong spin-orbit coupling. Here, we report on stacking-order effect of spin-orbit torque (SOT), spin-Hall magnetoresistance, and magnetic anisotropy in bilayer $Ni_{81}Fe_{19}$-5$d$ iridium oxide, $IrO_2$. While all the $IrO_2$ and Pt control samples exhibit large dampinglike-SOT generation stemming from the efficient charge to spin current conversion, the magnitude of the SOT is larger in the $IrO_2$ (Pt)-bottom sample than in the $IrO_2$ (Pt)-top one. The fieldlike-SOT has even more significant stack order effect, resulting in an opposite sign in the $IrO_2$ samples in contrast to the same sign in the Pt samples. Furthermore, we observe that the magnetic anisotropy energy density and the anomalous Hall effect are increased in the $IrO_2$ (Pt)-bottom sample, suggesting enhanced interfacial perpendicular magnetic anisotropy. Our findings highlight the significant influence of the stack order on spin transport and magnetotransport properties of Ir oxide/ferromagnet systems, providing useful information on design of SOT devices including 5$d$ transition-metal oxides.

* kueda@phys.sci.osaka-u.ac.jp

# I. INTRODUCTION

Interfaces composed of ferromagnet (FM) and non-magnet (NM) with strong spin-orbit coupling (SOC) have been an important arena to explore spin-current-related phenomena. One of the interesting subjects is spin-orbit torque (SOT), which is generated by in-plane current through the NM|FM bilayers and subsequent spin accumulation through charge to spin-current conversion [2]. The relevant conversion mechanism can be spin Hall effect (SHE) of the bulk NM [3] and/or interface Rashba-Edelstein effect (REE) [4], while both effects are triggered by the strong SOC. Irrespective of these origins, the SOT is highly sensitive to interface properties since the spin accumulation occurs at interface; the SOT has been established as a powerful way to clarify the interface properties [2,5–22]. The NM|FM interfaces can be examined as well by other techniques such as spin Hall magnetoresistance (SMR) [23–25], spin pumping [26,27], spin torque ferromagnetic resonance [6], perpendicular magnetic anisotropy (PMA) [28–30], anomalous Hall effect (AHE) [31–36], and magnetic proximity effect [37–39]. The validity of these measurements has been widely demonstrated by using $5d$ transition metals as NM: Pt, Ta, and W (Refs. [5–39]).

A new class of spintronic materials, $5d$ transition-metal oxides, is of peculiar interest because of its unique electronic structure; the Fermi surface is dominated only by $5d$ electrons with strong SOC, in contrast to that of $5d$ transition metals dominated by both $5d$ and $6s$ electrons. Recently, spintronic studies related to conductive Ir oxides have been attracting great attention [40–48]; several research groups have experimentally proven sizable SOT via SHE in epitaxial $SrIrO_3$ [43–45], epitaxial $IrO_2$ [48], and amorphous $IrO_2$ [47]. These reports highlight efficient charge to spin-current conversion of the Ir oxides, suggesting that the unique electronic structure inherent to $5d$ orbital plays a crucial role in the spin-transport mechanism. While the Ir oxides thus have been a potential component for spintronic devices, there remains unsolved issues on device design when we combine the Ir oxides with FM. One of the most important issues is stack order of the bilayer structure, which greatly influences on the interface properties. In typical Pt|FM|Pt layers with two interfaces, for instance, one of the interfaces is indeed found to largely contribute to PMA [30,49,50] and SOT [19]. Thus, inverting the stack order in the bilayer is an effective approach allowing us to further understand and control interfacial-driven properties such as SOT, SMR, AHE, and PMA.

In this work, we report that the stack order of bilayers $Ni_{81}Fe_{19}(Py)|IrO_2$ strongly influences on spin transport, electrical and magnetic properties evaluated by SOT, SMR and magnetotransport measurements. We choose amorphous $IrO_2$ since it can be easily incorporated into devices due to its simplicity of deposition. The stack order affects the DL-SOT and the PMA. Their magnitudes are larger in $IrO_2$-bottom sample than $IrO_2$-top one, whereas the same behavior is observed in Pt control samples. The fieldlike (FL)-SOT has shown even more drastic stack order effect, resulting in an opposite sign between the IrO$_2$-bottom and IrO$_2$-top samples in contrast to the same sign in the Pt samples. These findings show up the significance of the stack order in designing spintronic devices including 5$d$ transition-metal oxides.

## II. EXPERIMENT

We grew two types of samples comprising of Sub.|1.5 TaO$_x$|3 Ni$_{81}$Fe$_{19}$|10 IrO$_2$ and Sub.|10 IrO$_2$|3 Ni$_{81}$Fe$_{19}$|2.3 TaO$_x$ on thermally oxidized Si substrates by magnetron sputtering [Fig. 1(a)]. These samples are labeled as IrO$_2$-top (T) and IrO$_2$-bottom (B). The number indicates the layer thickness in nanometer. The Ni$_{81}$Fe$_{19}$ alloy is a soft ferromagnet with in-plane magnetic anisotropy, referred as permalloy (Py). Before the deposition of the Py layer, we eliminate surface oxidation of the Py target with sufficient sputter time. The TaO$_x$ layers are obtained by deposition of Ta and subsequent oxidation from the SiO$_2$ substrate (IrO$_2$-T) and air (IrO$_2$-B); in this experimental design, we expect to prevent the oxidation of Py surfaces at a side opposite to IrO$_2$. Considering the previous study [18] and our sample structures, the TaO$_x$ layers would not influence interfacial roughness of the bilayer, suggesting that the layers are not crucial in spin transport and magneto transport effects. We expect that current shunting by the TaO$_x$ layers can be neglected due to their high resistance. The amorphous IrO$_2$ is grown by a reactive sputtering method at the rate of Ar:O$_2$ = 8:2 [47]. The surface of the Py layer is exposed to the Ar-O$_2$ gas only in the IrO$_2$-T sample, giving rise to additional oxidation of the Py layer as discussed later. We also prepare control samples including Pt, namely, Sub.|1.5 TaO$_x$|3 Ni$_{81}$Fe$_{19}$|6 Pt (Pt-T) and Sub.|6 Pt|3 Ni$_{81}$Fe$_{19}$|2.3 TaO$_x$ (Pt-B). All of the metal and oxide layers are grown at total deposition pressure of ~0.4 Pa, while no annealing treatment before or after deposition is provided.

The Hall bar structure, fabricated using photolithography and post-deposition lift-off, has channel dimensions of 10 μm width ($w$) and 30 μm length ($l$), as illustrated in Fig. 1(a). The $I$ indicates ac or dc current flows producing the current-driven SOT, which has two components with different symmetries, namely, dampinglike (DL) and fieldlike (FL) SOTs, corresponding to effective fields $B_{DL}$ and $B_{FL}$, respectively. These fields represent $B_{DL} \parallel \sigma \times M$ and $B_{FL} \parallel \sigma$, where the direction of accumulated spin $\sigma$ at interface is along $y$. The Hall bars were mounted on a motorized stage allowing for in-plane ($\phi$) and out-of-plane ($\theta$) rotation, with angle and coordinate system in Fig. 1(a). The samples were placed in an electromagnet producing fields in the range of 0.05 to 1.5 T. The experiments were performed at room temperature using ac current $I_{ac}=\sqrt{2}I_{rms}\sin(2\pi ft)$ with $f$ =13 Hz for resistivity, anomalous Hall and harmonic Hall measurements and using dc current $I_{dc}$ for magnetoresistance measurement. We set $I_{rms}$=50 μA for the anomalous Hall and $I_{rms}$=1–2 mA for harmonic Hall measurements, and $I_{dc}$=500 μA for magnetoresistance measurement.

## III. RESULTS

### A. Electrical transport and magnetic properties

We examined the saturation magnetization ($M_s$) of each bilayer by measuring in-plane hysteresis loop using a superconducting quantum interference device magnetometer. Figure 1(b) shows the $M_s$ for all the samples. We find that the $M_s$ values of the Pt samples are larger than those of the IrO$_2$ samples; we estimate $M_s$ [Pt] of 5.80–6.05 × 10$^5$ A/m and $M_s$ [IrO$_2$] of 4.90–5.35 × 10$^5$ A/m. We associate the reduction in the IrO$_2$ samples with partial oxidation of the Py layer. In particular, the reduction of $M_s$ is larger in the IrO$_2$-T sample than in the IrO$_2$-B sample. Compared with the Pt samples, the reduction is roughly 20% in the IrO$_2$-T sample. This corresponds to the oxidized and probably nonmagnetic Py thickness of 0.6 nm, while the thickness is estimated to be 0.2 nm for the IrO$_2$-B sample. Figure 1(c) displays the sheet resistance ($R_{sq}$) for all the samples. We observe a higher $R_{sq}$ in the IrO$_2$ samples than that in the Pt samples resulting from the more resistive IrO$_2$ than Pt. By focusing on the stack order, the $R_{sq}$ in the IrO$_2$-T sample is 14% larger than that in the IrO$_2$-B one, while the $R_{sq}$ is almost the same between the Pt-T and Pt-B samples. This difference comes from the oxidation effect in the IrO$_2$-T sample introduced by the above-mentioned difference of the deposition condition depending on the stack order; the oxidation of the Py surface in the IrO$_2$-T sample is larger than that in the IrO$_2$-B one. In order to discuss the oxidation effect, we calculate the resistivity ($\rho$) within a parallel resistor model. According to the $\rho$ (IrO$_2$) ~ 370 μΩcm and $\rho$ (Pt) ~ 26 μΩcm from our previous work [47] and the measured $R_{sq}$, we extracted the $\rho$ (Py) ~ 105 μΩcm for the IrO$_2$-T sample, $\rho$ (Py) ~ 87 μΩcm for the IrO$_2$-B one and $\rho$ (Py) ~ 83 μΩcm for the Pt samples. The resistive Py in the IrO$_2$ samples result from the reduced Py thickness due to oxidation. Assuming that the oxidized Py is fully insulating, thickness of the oxidized Py layer is calculated to be ~0.6 nm and ~0.2 nm for the IrO$_2$-T and IrO$_2$-B samples, respectively. These values are consistent with the ones obtained from $M_s$ and therefore we consider that Py in the IrO$_2$-T sample is significantly more oxidized than Py in the IrO$_2$-B one.

We measure the anomalous Hall resistance ($R_{AHE}$) and effective anisotropy field ($B_k^{eff}$) by sweeping a perpendicular magnetic field ($B_z$) at $\theta = 0°$ to determine the perpendicular magnetic anisotropy energy density ($K$) by using the following relation: $K = \frac{M_s}{2}(\mu_0 M_s - B_k^{eff})$ [51]. Figure 2(a) shows the representative result of the IrO$_2$-T sample. Since the first harmonic Hall resistance ($R_H^{1\omega}$) is saturated in the high field regime, we obtain $R_{AHE}$ from subtraction between two extrapolated high-field slopes (black lines). The $B_k^{eff}$ is estimated from subtraction between two intersections (black dotted line) given by a linear fit of the low field data with black line. The $K$ for all the samples is displayed in Fig. 2(b). The $K$ of the IrO$_2$ (Pt)-B sample is found to be significantly increased by a factor of about 6 compared to the corresponding the IrO$_2$ (Pt)-

T sample. Since the bottom Pt layer is known to play a dominant role in the strong PMA in the typical Pt|FM bilayers [10,12,16,18,49,52], we relate the substantial difference to be enhanced interface-driven PMA in the Pt-B sample, resulting in large $K$. The correlation between PMA and contribution of orbital moment has been experimentally found [49,52]; strong PMA with large orbital moment is induced when FM on Pt is prepared in contrary to FM underneath Pt [49]. We then focus on Fig. 2(c) that displays the $R_\mathrm{AHE}$ for all the samples. Similar to the trend of $K$, $R_\mathrm{AHE}$ of the $IrO_2$ (Pt)-B sample is increased by about 40 % than that of the $IrO_2$ (Pt)-T one. We point out that the AHE has been known to consist of bulk and interface contributions [31–39, 53–56]. The former is the resistivity effect [36], while the latter includes several origins such as the magnetic proximity effect [37–39], intermixing [53] and the interface SOC [31,54–56]; the SOC also plays a crucial role in the strong PMA [31,54–56]. The observed enhancement of AHE in the $IrO_2$ (Pt)-B sample cannot be explained by the resistivity effect [36] since the $IrO_2$-B sample has lower $R_\mathrm{sq}$ than that in the $IrO_2$-T one [see Fig. 1(c)]. Considering that the correlation of $K$ and $R_\mathrm{AHE}$ for the $IrO_2$ (Pt)-B sample is obvious in Fig. 2(b)-2(c), our data indicates the large interface contribution at the $IrO_2$ (Pt)-B sample. Whereas such interface-driven AHE and PMA in amorphous $IrO_2$-FM system remains unclear at this stage, the contribution of orbital moment is likely in common with Pt considering the observed similarity between the $IrO_2$ and Pt samples.

**B. Magnetoresistance measurement**

We present the description of magnetoresistance measurement. The first harmonic longitudinal resistance ($R_\mathrm{L}^{1\omega}$) that is equivalent to standard dc measurement can be typically expressed by the general form [14]:

$$R_\mathrm{L}^{1\omega} = R_0 - \Delta R_\mathrm{zx} \sin^2\theta \cos^2\phi + \Delta R_\mathrm{zy} \sin^2\theta \sin^2\phi, \quad (1)$$

where $R_0 \equiv R(M\|x)$, $\Delta R_\mathrm{zx}$ is the resistance difference when the $M$ sufficiently saturates along $z$ axis and $x$ axis directions with $B_\mathrm{ext}$, $\Delta R_\mathrm{zy}$ is the resistance difference when the $M$ sufficiently saturates along $z$ axis and $y$ axis directions with $B_\mathrm{ext}$. We also define $\Delta R_\mathrm{xy}$, which is the resistance difference between $M$ at $x$ and $y$ directions, corresponding to $\Delta R_\mathrm{zy} - \Delta R_\mathrm{zx}$.

We measured the $R_\mathrm{L}^{1\omega}$ by rotating a sample with a fixed $B_\mathrm{ext} = 1.35$ T in three orthogonal planes, shown in Fig. 3(a). The applied $B_\mathrm{ext}$ is large enough to saturate the $M$ along all coordinate axis, allowing us to characterize magnetoresistances such as $\Delta R_\mathrm{zy}$, $\Delta R_\mathrm{xy}$ and $\Delta R_\mathrm{zx}$. Examining in these planes is essential to distinguish the contribution of SMR and AMR. The $zy$ scan illustrates the SMR, which is magnetoresistance due to asymmetry between absorption and reflection of spin current generated from the bulk SHE in NM layer [23–25]. Accordingly, this contribution gives higher resistance at $M//z$ and lower resistance at $M//y$, resulting in $\Delta R_\mathrm{zy} \approx$ $\Delta R_{xy} > 0$ and $\Delta R_{zx} \approx 0$. The *zx* scan illustrates the AMR, originating from the enhanced scattering of conduction electrons from the localized *d*- orbitals (*s-d* scattering) in the bulk FM [57]. This contribution gives the higher resistance at *M*//*x* and lower resistance at *M*//*z*, resulting in $\Delta R_{zx} \approx \Delta R_{xy} > 0$ and $\Delta R_{zy} \approx 0$. The *xy* scan includes both SMR and AMR contributions in NM contacted with conducting ferromagnet. Figure 3(b) shows the representative magnetoresistance of the IrO$_2$-T sample for three planes. In order to determine the MR ratio, the amplitudes ($\Delta R_{zy}$, $\Delta R_{zx}$, and $\Delta R_{xy}$) were independently obtained by the fit using Eq. (1) on the respective MR curves. We confirmed that the relationship $\Delta R_{xy} = \Delta R_{zy} - \Delta R_{zx}$ within the error bars, indicating that our fit is valid to determine each amplitude of the MR ratio. Note that the slight discrepancy between the data and fit in the *zy* and *zx* plans is observed, possibly from a contribution of higher order MR stemming from texture effect [58].

Figures 3(c)-(e) summarize normalized magnetoresistance for all the samples defined as $\Delta R_{zy,zx,xy}/R_0$. The largest magnetoresistance appears in *xy* planes for all the samples, resulting from the combination of SMR and AMR. Here, we point out recent MR studies in bilayers that have been a subject of ongoing debate [58–63]. The possible contributions are the anisotropic interface MR [58,59], the anomalous Hall MR [62], the absorption of spin current in FM layer [25,61], and spin current generation from FM layer [60,61,63]. Considering the above effects, the use of conducting FM makes the SMR effect more complicated compared to the SMR observed in Pt contacted with magnetic insulators [23]. The magnitude of the $\Delta R_{zy}/R_0$ is larger than that of the $\Delta R_{zx}/R_0$ indicating that the SMR contribution is dominant in all the samples; we expect that the $\Delta R_{zy}/R_0$ mostly includes contribution of the SMR discussed in Sec. III. D. The larger SMR for the Pt than the IrO$_2$ samples is observed as well. This is ascribed to current shunting effect because of more current flow in Pt than IrO$_2$. We also find that both SMR and AMR are enhanced in the IrO$_2$ (Pt)-bottom samples. If the $\Delta R_{zx}/R_0$ is fully contributed from AMR of the bulk FM, the stack order effect should be opposite according to the current shunting effect. While this strongly suggests that the observed $\Delta R_{zx}/R_0$ includes some interface contribution, we do not discuss the $\Delta R_{zx}$ in detail since it is difficult to pin down a role of the interface at current stage. However, given the report on the MR in bilayer including IrO$_2$, it serves as a good starting point to clarify complicated MR with further study. Unlike the $\Delta R_{zx}$, the $\Delta R_{zy}$ is well related to DL-SOT efficiency; by analyzing the efficiency through conventional SMR model, we elucidate the stack order effect in Sec. III. D.

**C. Harmonic Hall measurement**

We quantify DL-SOT and FL-SOT by performing harmonic Hall measurement described in Ref. [9–11,14]. While the $R_H^{1\omega}$ is equivalent to conventional Hall resistance, the second harmonic resistance $R_H^{2\omega}$ provides information about the SOT; an injection of $I_{ac}$ produces

SOTs that cause the small modulation of the magnetization about its equilibrium position against magnetic field. In the analysis of the harmonic Hall measurement established by Avci [11], the following relation is used when magnetization sufficiently lies in-plane ($\theta = 90°$):

$$R_H^{1\omega} = R_{PHE} \sin 2\phi \sin^2\theta \quad (2).$$

$$R_H^{2\omega} = -\left(R_{AHE} \frac{B_{DL}}{B_k^{eff}+B_{ext}} + R_{\nabla T}\right)\cos\phi + 2R_{PHE}\frac{B_{FL}+B_{Oe}}{B_{ext}}(2\cos^3\phi - \cos\phi). \quad (3)$$

$$\equiv R_{DL+\nabla T} \cos\phi + R_{FL+Oe}(2\cos^3\phi - \cos\phi). \quad (4)$$

Here, $R_{PHE}$, $R_{\nabla T}$, and $B_{Oe}$ are the planar Hall resistance, the thermal induced second harmonic resistance, and the current induced Oersted field, respectively. Since both the DL and the thermal induced contributions have the same symmetry, they appear in pairs in Eq. (3); it is also the case for the FL and Oersted field contributions. In this case, as indicated in Eq. (4), we define $R_{DL+\nabla T}$ and $R_{FL+Oe}$ as the coefficients of the $\cos\phi$ and $(2\cos^3\phi - \cos\phi)$ components, which can be separated by the fitting on $R_H^{2\omega}$ versus $\phi$. The most convenient way to extract all the relevant parameters is to measure both $R_H^{1\omega}$ and $R_H^{2\omega}$ as functions of $\phi$ with various $B_{ext}$.

Following the above procedure, we extract $B_{DL}$ and $B_{FL}$. Representative data of $R_H^{1\omega}$ and $R_H^{2\omega}$ with $B_{ext}$ = 0.1 and 0.5 T in the IrO$_2$-T sample are exemplified in Fig. 4(a) and 4(b) [top panel], respectively. The amplitude of $R_H^{1\omega}$ is independent of the $B_{ext}$, indicating that the $R_{PHE}$ is fully saturated magnetization in in-plane $xy$ axis at 0.1 T; $R_{PHE}$ = 0.12 Ω was obtained as the fitting results (black curves) in accordance with Eq. (2). On the other hand, $R_H^{2\omega}$ is strongly field-dependent as expected from Eq. (3), reflecting the modulation of SOT. In order to independently extract $R_{DL+\nabla T}$ and $R_{FL+Oe}$, we fit $R_H^{2\omega}$ by using Eq. (4) to plot components of $\cos\phi$ and $(2\cos^3\phi - \cos\phi)$, as DL- (middle panel) and FL-contributions (bottom panel), respectively. These coefficients are plotted as a function of $1/(B_{ext} + B_k^{eff})$ and $1/B_{ext}$, shown in Figs. 4(c) and 4(d). The slopes of these curves correspond to $R_{AHE}B_{DL}$ and $2R_{PHE}B_{FL+Oe}$, respectively, from which $B_{DL}$ and $B_{FL+Oe}$ are extracted. In order to estimate the $B_{FL}$, $B_{Oe}$ is subtracted from $B_{FL+Oe}$ through the Ampere's law as $B_{Oe} = \mu_0 J t_{NM}/2$, where $t_{NM}$ is the non-magnet layer thickness and the $J$ is the current density flowing in NM layer given by $J = I/(t_{NM}w)$. We exclude two points in low field regime deviating from the linear relation since the $M$ is not fully saturated in the $xy$ plane; the $B_{ext}$ required to saturate the moments is larger than the shape anisotropy field defined by the geometry of the Hall bar [64]. Considering that all the samples exhibit the similar behavior, the deviation is probably due to pinning effect, irrespective of the oxidation of the Py layer. The intercept of Fig. 4(c) indicates the thermal effect $R_{\nabla T}$ is caused by an increase in Joule heating [11,14,47] due to the higher resistance of the IrO$_2$ samples; the Pt samples exhibit smaller $R_{\nabla T}$ due to their lower resistance. With the parallel resistor model, we determine the current fractions to be 45–50 % for the IrO$_2$ layers and ~88 % for the Pt layers using each resistivity of IrO$_2$, Pt and Py discussed in Sec. III. A.

Figs. 4(e) and 4(f) display $B_{DL}/J$ and $B_{FL}/J$ for all the samples, normalized at $J=10^{11}$ A/m$^2$, respectively. We find sizable SOT effective fields for all the samples, which allow us to evaluate their SOT efficiencies. The sign of DL-SOT is positive for all the samples, whereas that of FL- SOT is positive for the Pt samples but opposite between the IrO$_2$ samples: negative for the IrO$_2$-T and positive for the IrO$_2$-B. The magnitude of the SOT efficiency and the sign change of FL-SOT are discussed in next section.

**D. Discussion of spin-orbit torque efficiency**

The SOT efficiency observed by the two methods (SMR and harmonic Hall) provides an insight to understand the stack order effect of spin-transport properties. The SMR method leads to the DL-SOT efficiency ($\xi_{DL}$), while the harmonic Hall method provides both the $\xi_{DL}$ and FL-SOT efficiency ($\xi_{FL}$). Based on the drift-diffusion model assuming perfect interface transparency, the ratio of SMR can be related to the DL-SOT efficiency as follows: [24]

$$\Delta R_{zy}/R_0 = \xi_{DL}^2 \frac{\lambda}{t_{NM}} \frac{\tanh(t_{NM}/2\lambda)}{1+a}\left[1-\frac{1}{\cosh(t_{NM}/\lambda)}\right], (5)$$

where $\lambda$ is spin diffusion length of the NM layer and $a$ is the current shunting defined as $a \equiv \rho_{NM}t_{FM}/\rho_{FM}t_{NM}$ due to the presence of the conducting FM layer. The $t_{FM}$ indicates the magnet layer thickness. Here, we use $\lambda = 1.7$ nm for IrO$_2$ [47] and $\lambda = 1.4$ nm for Pt [16]. The $B_{DL}$ and $B_{FL}$ obtained from harmonic Hall measurement can be converted to $\xi_{DL}$ and $\xi_{FL}$ according to the following expression [65]

$$\xi_{DL(FL)} = \frac{2eM_S t_{FM}}{\hbar}\frac{B_{DL(FL)}}{J}, (6)$$

where $\hbar$ is the Dirac constant and $e$ is the elementary charge. The results of SMR and harmonic Hall methods for all the samples are summarized in Figs. 5(a) and 5(b).

First, we discuss $\xi_{DL}$ determined from both harmonic Hall and SMR methods. In Fig. 5(a), the data exhibits the similar trend on both methods, while the SMR data is smaller than the harmonic Hall data; the difference ranges from 8 to 17 %. This discrepancy is attributed to the other contributions on the observed SMR discussed in Sec. III. B. However, qualitative agreement of the $\xi_{DL}$ measured by both methods clearly shows that the $\xi_{DL}$ is mainly from SMR contribution. We hereafter focus on the harmonic Hall data for further discussion. As displayed in Fig. 5(a), the harmonic Hall data from 0.076 to 0.100 with positive sign for all the samples, which is close to the conventional value of ~0.1 for IrO$_2$ [47] and Pt [5,6,10,11,14,16,18]; we consider that the bulk SHE is origin of the DL-SOT in similar to the previous studies [6,10,11,14–16,18,47]. This result indicates that the interface of Py and IrO$_2$ plays a significant role in efficient SOT generation.

Second, we find that the stack order greatly influences on the $\xi_{DL}$ in Fig. 5(a). The magnitude of the IrO$_2$ (Pt)-B sample is found to be about 22% (15%) larger than that of the IrO$_2$ (Pt)-T one. Previous work in Py-Pt bilayers has shown the similar stacking order effect on the SOT, which strongly supports our observation [66]. This is attributed to reduced spin transparency in the NM-T samples since low spin transparency at interface typically provides the small SOT generation [15,16]. We point out that the situation between the Pt and IrO$_2$ samples is different in terms of resistance; the Pt samples have the same $R_{sq}$ while the IrO$_2$-T sample have much higher resistance than IrO$_2$-B sample due to the oxidized Py layer as mentioned in Sec. III A. Considering that the variation of the $\xi_{DL}$ is nearly the same in the IrO$_2$ and Pt samples, the oxidation effect is not a main factor in reducing DL-SOT, although it is crucial to increase $R_{sq}$. A possible factor is the presence of intermixing between FM and NM layers, which could lead to the formation of a nonmagnetic or weakly magnetic surface alloy [12,22,50]. In Pt|Co bilayers by inserting ultra-thin spacer layers at the interface, the reduction of DL-SOT is more drastic than the increase of $R_{sq}$, suggesting that the intermixing effect strongly influences on SOT property rather than sample resistance does [22]. Moreover, there are some reports that the deposition of NM on FM provides small contributions of interface properties such as PMA and SOT [19,30,49,50]. Thus, the intermixing effect is likely to be larger in the top NM rather than the bottom NM contacted with FM as well in our study although it is hard to separate the intermixing from oxidation effects in the IrO$_2$ samples. Following this, the IrO$_2$ (Pt)-T sample exhibits low spin transparency at interface, resulting in the reduction of $\xi_{DL}$. Our experimental result therefore demonstrates a role of intermixing effect for the DL-SOT generation in the IrO$_2$ samples by comparing with the Pt control samples. Here we note that a recent spin pumping experiment reports on an enhancement of the Gilbert damping due to interfacial oxidation in SIO|Py bilayers [67], indicating that spin current transport might be modified as well. Our observation will be a step to further understand spin-current physics in the 5$d$ transition-metal oxides.

Finally, we discuss the FL-SOT in Fig. 5(b). The absolute value of $\xi_{FL}$ for all the samples, ranging from 0.010 to 0.024, is much smaller than the above mentioned $\xi_{DL}$, in good agreement with typical bilayers found in Refs. [11,16,18,47]. Of particular interest is the opposite sign between the IrO$_2$ samples. Previously, we have reported the negative sign with a possible contribution of interface REE in Py|IrO$_2$, which has the same structure with the IrO$_2$-T sample [47]; similar results are found in metallic bilayers [16,21,22]. Our observation in this study emphasizes that stack order is crucial on the determination of the sign. This suggests that the opposite sign is not related to the intermixing effect since the sign for the Pt samples remains the same. Note that if the bulk SHE is dominant source in the FL-SOT, the positive signs of both the $IrO_2$-B and Pt-B samples are reasonable, according to the direction of $\sigma$. Here, we attribute a possible explanation of the negative sign in the $IrO_2$-T sample to the oxidation effect introduced by inverting the stack order. The effect causes the formation of $PyO_x$ in Py surface, enhancing the interfacial REE found in Refs. [68–70]; a theoretical work also suggests that the REE is reinforced by oxidation [71]. If the $PyO_x$ layer as thick oxidation barrier is formed, the spin transparency is significantly decreased, resulting in less SOT generation [69,72]. Hence, we found that the oxidation effect plays a major role in FL-SOT in contrast to DL-SOT. By inverting the stack order, our experimental evidence provides a clue for manipulating the sign of FL-SOT, with preservation of large DL-SOT, that is, efficient charge to spin current conversion. Finally, we refer to the spin memory loss (SML) at the interface; many efforts so far have been made to quantify the SML in Pt/Py bilayers [66,73], suggesting that it would become another source for changes in the SOT efficiency. Further studies will be required to clarify a microscopic origin of the SOT such as spin pumping, spin diffusion, and SML.

**IV. CONCLUSION**

We investigated the stack order effect of spin-transport and magnetotransport properties by performing SOT, SMR, and AHE measurements in Py-$IrO_2$ bilayers. We found the significant stack order effect on the DL-SOT, the FL-SOT, and PMA. The DL-SOT in the $IrO_2$ (Pt)-B sample exhibits a larger magnitude than that in the $IrO_2$ (Pt)-T one, with positive sign for all the samples. The reduction of the $IrO_2$ (Pt)-T sample is attributed to be the intermixing effect, reducing the spin transparency at the interface; common behavior in the $IrO_2$ and Pt samples is observed. The FL-SOT has an opposite sign in the $IrO_2$ samples in contrast to the same sign in Pt samples, giving the positive sign for the $IrO_2$-B and Pt samples and the negative for the $IrO_2$-T sample. The negative sign of the FL-SOT stems from the interface REE, possibly enhanced by the formation of $PyO_x$ layer between Py and $IrO_2$ layers; the oxidation effect for the $IrO_2$-T sample plays a major role in the FL-SOT generation. Finally, we observe that both perpendicular magnetic anisotropy energy density and anomalous Hall effect are increased in the $IrO_2$-B sample, suggesting the enhancement of perpendicular anisotropy at interface, as similar results in typical Pt|FM. Thus, inverting the stack order significantly influences on interface-driven properties, offering alternative way to control the sign and magnitude of SOTs as well as magnetic anisotropy in amorphous $IrO_2$ based bilayers. These information is useful for developing the device design, leading to electrically controlled memory and logic devices using $5d$ transition-metal oxides|FM systems.

ACKNOWLEDGEMENT

The authors thank T. Arakawa for technical support. This work was carried out at the Center for Advanced High Magnetic Field Science in Osaka University under the Visiting Researcher's Program of the Institute for Solid State Physics, the University of Tokyo. This work was partly supported by Japan Society for the Promotion of Science Grant-in-Aid for Young Scientists (Grant No. 19K15434), Scientific Research on Innovative Areas "Quantum Liquid Crystals" (KAKENHI Grant No. JP19H05823), JPMJCR1901 (JST-CREST), and Nippon Sheet Glass foundation for Materials Science and Engineering. We acknowledge the stimulated discussion in the meeting of the Cooperative Research Project of the Research Institute of Electrical Communication, Tohoku University.

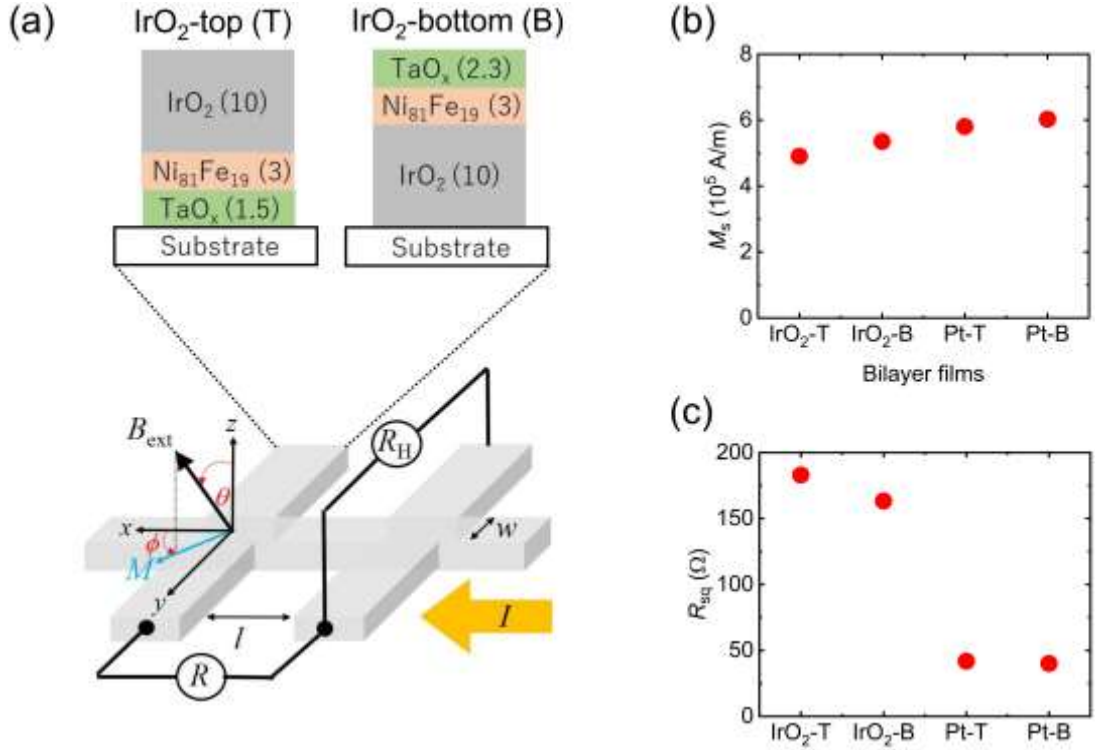

FIG. 1. (a) Top: Cross section of the bilayer films labeled as IrO$_2$-top (T): 3 Py|10 IrO$_2$ and IrO$_2$-bottom (B): 10 IrO$_2$|3 Py. Bottom: Patterned device structure with two Hall bars, electrical detection and coordinate system. (b) Saturation magnetization and (c) sheet resistance for the samples with IrO$_2$-T and IrO$_2$-B, including Pt control samples of 3 Py|6 Pt and 6 Pt|3 Py.

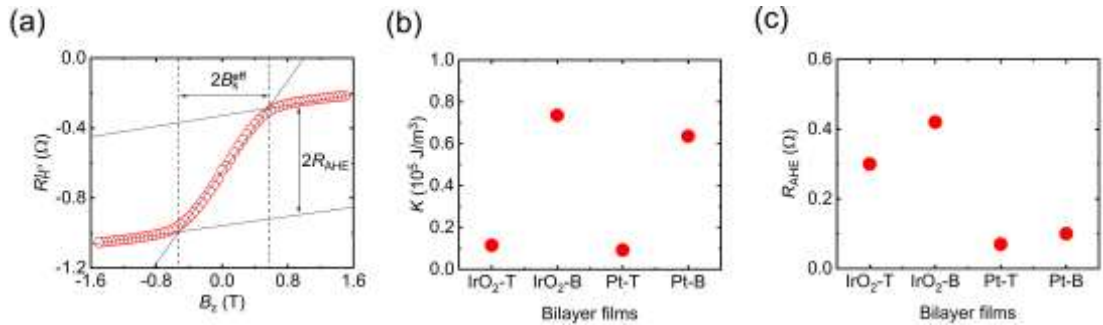

FIG. 2. (a) Hall resistance of IrO$_2$-T sample as a function of perpendicular magnetic field at $\theta = 0°$. (b) Perpendicular magnetic anisotropy energy density and (c) anomalous Hall resistance for all the samples extracted from measurements similar to the one shown in (a).

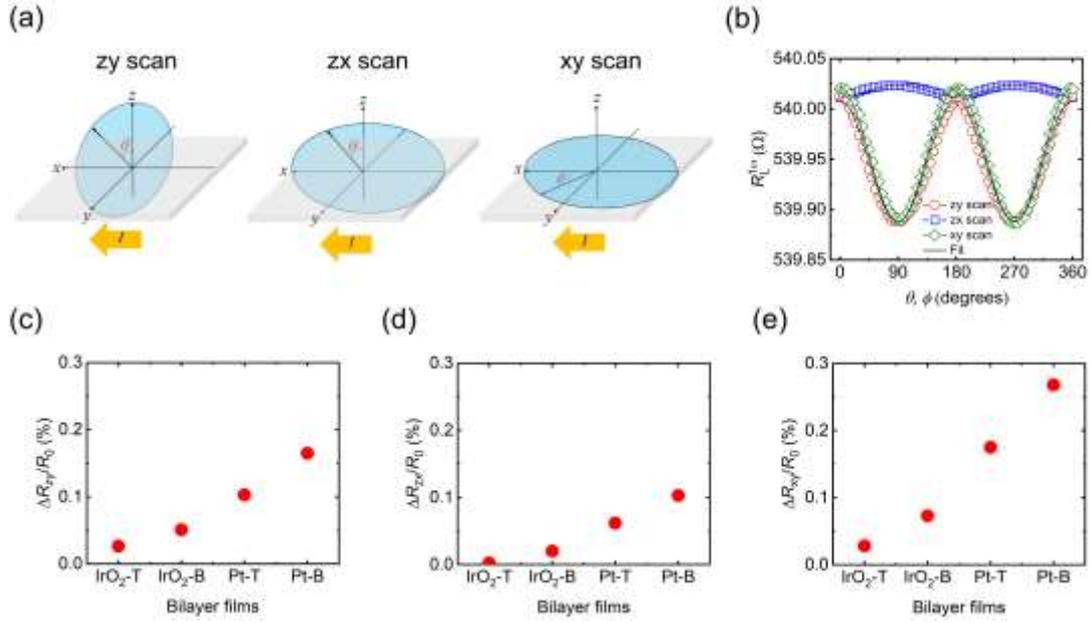

FIG. 3. (a) Illustration of the rotation planes. (b) First-harmonic longitudinal resistance of IrO$_2$-T sample measured by rotating the samples with a fixed external field of 1.3 T. Black line is the result on fit using Eq. (1). (c-e) Magnetoresistance expressed in the percentage in three planes for all the samples.
18

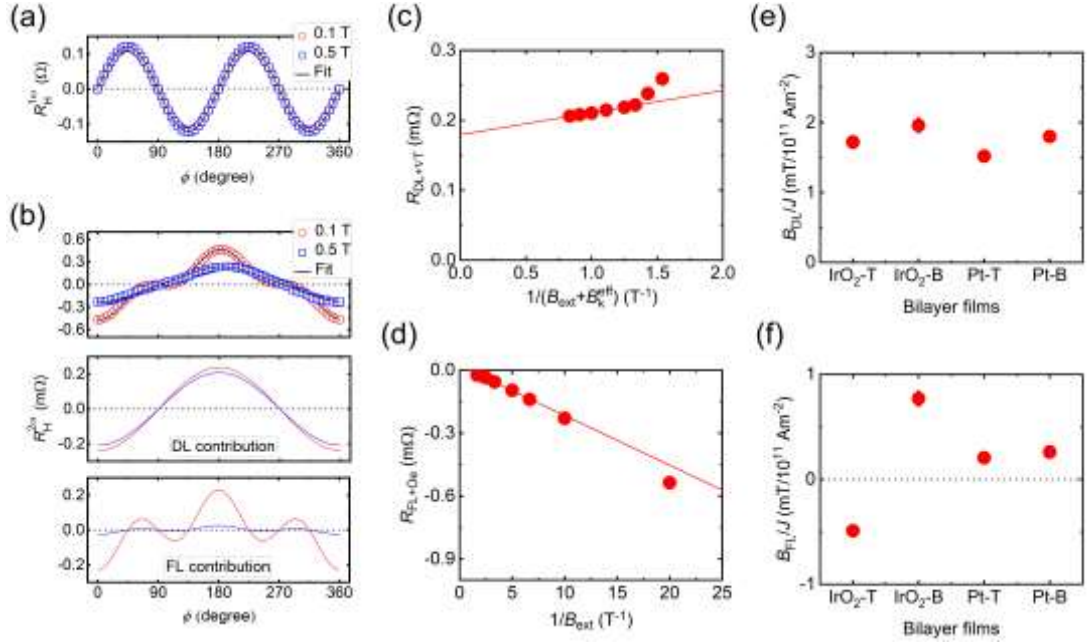

FIG. 4. (a) First-harmonic Hall resistance ($R_H^{1\omega}$) of $IrO_2$-T sample measured at 0.1 T (red open circle) and 0.5 T (blue open square). The black solid curve is fit to the data using Eq. (2). (b) Second-harmonic Hall resistance ($R_H^{2\omega}$) of $IrO_2$-T sample measured at 0.1 T and 0.5 T for raw data (top panel). The solid curves are fits to the data using Eq. (4). Separated $\cos\phi$ and ($2\cos^3\phi - \cos\phi$) components from $R_H^{2\omega}$ indicate DL contribution and FL contribution, respectively. (c) Obtained $R_{DL+\nabla T}$ as a function of $1/(B_{ext}+B_k^{eff})$ and (d) $R_{FL+Oe}$ as a function of $1/B_{ext}$ for the $IrO_2$-T sample. The red solid lines represent linear fits. (e) $B_{DL}/J$ and (f) $B_{FL}/J$ for all the samples. $J$ is the current density flowing in non-magnet layer normalized to $10^{11}$ $A/m^2$.

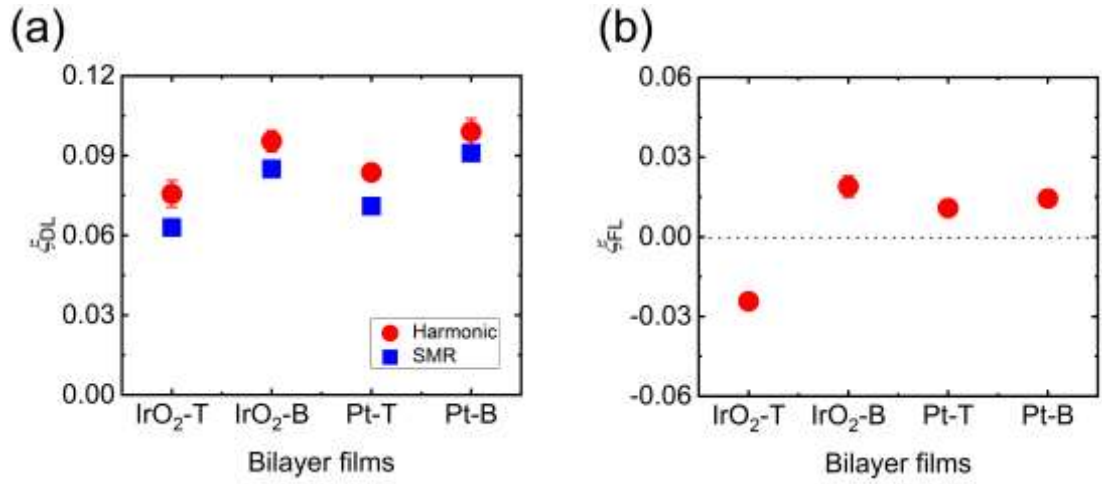

FIG. 5. (a) Dampinglike spin-orbit torque efficiency evaluated by harmonic Hall and spin-Hall magnetoresistance measurements for all the samples. (b) Fieldlike spin-orbit torque efficiency for all the samples.